# Topological Receptive Field Model for Human Retinotopic Mapping

Yanshuai Tu[1][0000-0002-4619-2613], Duyan Ta[1],
Zhong-Lin Lu[2,3][0000-0002-7295-727X] and Yalin Wang[1] [0000-0002-6241-735X] *

[1] Arizona State University, Tempe AZ 85201, USA
[2] New York University, New York, NY
[3] NYU Shanghai, Shanghai, China
* ylwang@asu.edu

**Abstract.** The mapping between visual inputs on the retina and neuronal activations in the visual cortex, i.e., retinotopic map, is an essential topic in vision science and neuroscience. Human retinotopic maps can be revealed by analyzing the functional magnetic resonance imaging (fMRI) signal responses to designed visual stimuli *in vivo*. Neurophysiology studies summarized that visual areas are topological (i.e., nearby neurons have receptive fields at nearby locations in the image). However, conventional fMRI-based analyses frequently generate non-topological results because they process fMRI signals on a voxel-wise basis, without considering the neighbor relations on the surface. Here we propose a topological receptive field (tRF) model which imposes the topological condition when decoding retinotopic fMRI signals. More specifically, we parametrized the cortical surface to a unit disk, characterized the topological condition by tRF, and employed an efficient scheme to solve the tRF model. We tested our framework on both synthetic and human fMRI data. Experimental results showed that the tRF model could remove the topological violations, improve model explaining power, and generate biologically plausible retinotopic maps. The proposed framework is general and can be applied to other sensory maps.

## 1 Introduction

It is of great interest to quantify, simulate, and understand the relation between the visual inputs and the neuronal response with the retinotopic maps [1, 2]. A precise retinotopic map provides opportunities to understand or even simulate various aspects of the visual system. For instance, the complex-log [3] model, which depicted a rough position map between the retina and the primal visual cortex (V1), explained the dynamics of spiral visual illusions [4]. The retinotopic maps discovered some stable visual regions [5]. Additionally, retinotopic map research holds great promise in understanding brain plasticity and improving rehabilitation's efficacy from visual impairments. Further, retinotopic maps have been clinically adopted to monitor the progress and recovery under amblyopia treatment [6], a disorder that affects about 2% of children and may cause significant visual impairment if untreated.

†The work was supported in part by NIH (R21AG065942, RF1AG051710 and R01EB025032) and Arizona Alzheimer Consortium

Typically, human retinotopic maps are obtained *in vivo* by analyzing cortical functional magnetic resonance imaging (fMRI) response signals to visual stimuli [7, 8]. Since the first fMRI work on retinotopic mapping, several analysis models were proposed [9–11] to decode perception parameters from the noisy fMRI signals. Among them, the population receptive field (pRF) model [10] generates state-of-the-art results and becomes a cornerstone in fMRI signal analysis of retinotopic maps [12].

Although the pRF model achieved great success, the pRF results are usually not topological. The topological condition is an essential requirement of retinotopic maps since neurophysiology studies have revealed nearby neurons have receptive fields at nearby locations in the image [13, 14] (the topological condition). The topological condition is also the requirement of the vision system's hierarchical organization [1]: each visual area represents a unique map of a portion of retina. If there are duplicated representations in a visual area, this visual area should be further divided into more areas. Besides, it is challenging to infer accurate visual-related quantification without a topological retinotopic map, e.g., visual boundaries, cortical magnification factor, visual anisomery. Therefore, topological retinotopic map results are vital for neuroscience research.

A variety of methods were developed to reduce topological violations in the post-processing of pRF results [15–20]. For instance, the model fitting [15] is widely used to fit the decoded parameters with an algebraic model for V1-V3. Smoothing methods, e.g., Laplacian smoothing [16], process visual parameters one by one but cannot ensure the topological condition. Surface registration of retinotopic mesh is also developed for post-processing of pRF results [18, 21]. To our knowledge, however, *none* of the existing methods have considered the topological condition when decoding fMRI signals.

It is advantageous but challenging to impose the topological condition when decoding fMRI. First, the topological condition is different across visual areas, while the precise visual areas are delineated upon the decoding results. One may think the segmentation is available through the anatomical surface. Unfortunately, all anatomical-based segmentation is not accurate [21], limited by its modality. Second, since the fMRI signal is noisy, it is challenging to segment visual areas from the noisy decoded results.

We propose a topological receptive field (tRF) framework, which imposes the topological conditions by integrating the topology-preserving segmentation and topological fMRI decoding into a coherent system. More specifically, we first segmented the visual areas by preserving the prior connectivity pattern of visual areas based on initial decoding and then modeled the fMRI decoding with the topological condition within each visual area. Since the new fMRI decoding results can refine the segmentation, we repeated the segmentation and decoding until theoretically guaranteed convergence.

We validated the proposed framework on both synthetic data and human retinotopy data. Our experiments showed the superiority of tRF over other state-of-the-art methods, including the pRF model, post-processing by model fitting, smoothing, or registration methods. To our knowledge, (1) it is the first work that enforces the topological condition in decoding retinotopic fMRI signals; (2) it is also the first automatic visual area segmentation with the topology organization preserved. Our framework is general and can be extended to other sensory maps since most of the human perception maps are spatially organized, e.g., auditory map [22] is spatially related to sound frequency.

## 2 Method

### 2.1 Retinotopic mapping experiment

We begin with a brief introduction to the retinotopic mapping experiment. Such an experiment acquires both structural MRI and fMRI signals, as illustrated in **Fig.** 1. The structural MRI (**Fig.** 1g) is used to reconstruct the cortical surface. The fMRI (**Fig.** 1b) is acquired to monitor neuron activities during the visual stimuli (**Fig.** 1a). The visual stimulation is carefully designed to encode the visual space (**Fig.** 1f) uniquely. Before the analysis, the fMRI volumes are preprocessed (**Fig.** 1c) and projected onto the cortical surface (**Fig.** 1h). Eventually, for each subject, there is a high-quality anatomical/structural cortical surface in the form of discrete manifold $S = (V, E, F)$ (where $V$ is the surface vertex set, $E$ is the edge set, and $F$ is the face set), together with fMRI time series $y_i(t)$ for each vertex $V_i \in V$ on the cortical surface.

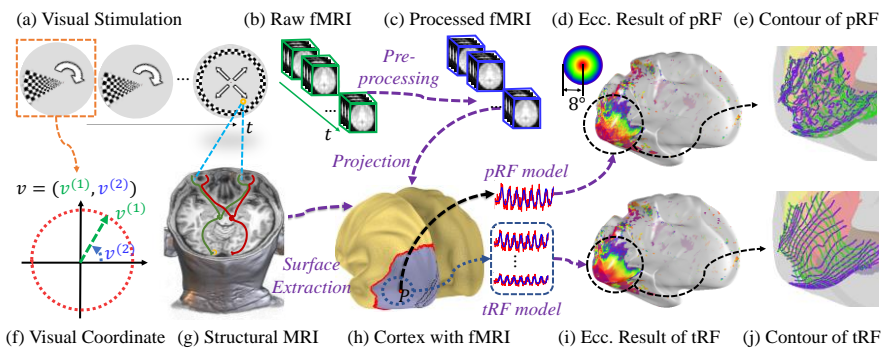

**Fig. 1.** The retinotopic map experiment and the pipelines of pRF/tRF models.

### 2.2 pRF model

The population receptive field (pRF) model decodes fMRI signals vertex by vertex. Namely, on the measurement $y_i(t)$, the pRF predicts perception parameters, including the population perception center $\boldsymbol{v_i} = \left(v_i^{(1)}, v_i^{(2)}\right)$ and population receptive field size $\sigma_i$. We take the polar coordinates for the visual fields, as shown in (**Fig.** 1f). Namely, $v_i^{(1)}$ and $v_i^{(2)}$ are the eccentricity and polar angle, respectively. Given the receptive model $r(\boldsymbol{v}'; \boldsymbol{v_i}, \sigma)$ [10], and the hemodynamic model $h(t)$ [23], the predicted fMRI is, $\hat{y}(t; \boldsymbol{v_i}, \sigma) = \beta(\int r(\boldsymbol{v}'; \boldsymbol{v_i}, \sigma)s(t, \boldsymbol{v}')d\boldsymbol{v}') * h(t)$, where $\beta$ is the activation level (a time-independent scalar). The perception center $\boldsymbol{v_i}$ and population receptive field size $\sigma_i$ are estimated by minimizing the error between the measured and predicted fMRI activation signal,

$$(\boldsymbol{v_i}, \sigma_i) = \arg \min_{(\boldsymbol{v_i}, \sigma_i)} E_p = \int \left(\hat{y}(t; \boldsymbol{v_i}, \sigma_i) - y_i(t)\right)^2 dt. \quad (1)$$

Solving **Eq.** 1 for all points generates the pRF retinotopic map. We show a typical pRF retinotopic map on the visual cortex in **Fig.** 1d. Unfortunately, a large portion of the result is not topological (the visual coordinates' contour curves are messed up in **Fig.** 1e). We are motivated to propose the tRF (**Figs.** 1ij) to improve the retinotopic maps.

## 2.3 Parametrization

To simplify the discussion, we first establish a parametrization between the 3D visual cortex and the 2D planar disk. We cut a geodesic patch containing the visual areas. In specific, we picked a point $p_0$ on the cortex, then enclosed a patch $P$ consist of cortical points within a certain geodesic to $P_0$, as illustrated in **Fig.** 2b. Then we computed the conformal parametrization for patch $P$ to planar disk $D$. If $c: P \to D$ is the conformal parameterization, then $u_i = c(V_i)$ where $u_i \in \mathbb{R}^2$ is the parametric coordinate for $V_i$. The conformal mapping was done via spherical conformal mapping of the double coverings of surface patches, followed by stereographic projection [24]. With the parametrization $c$, we can discuss the topology condition within the planar domains.

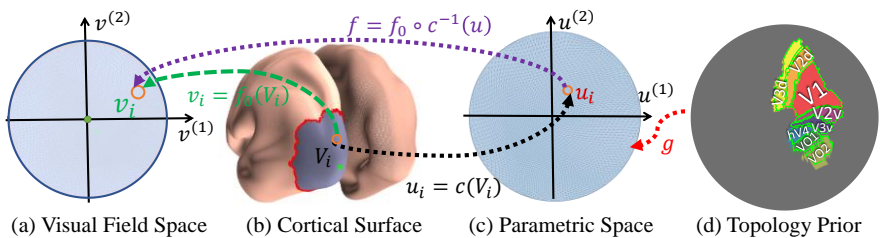

(a) Visual Field Space  (b) Cortical Surface  (c) Parametric Space  (d) Topology Prior

**Fig. 2.** Parametrization and segmentation: (a) the visual field, (b) the region of interest (ROI), (c) conformal parametrization of ROI, and (d) the topology-prior (the label is defined for each point).

## 2.4 Topology conditions

The topology conditions have two inferences. First, the visual coordinates are topological with respect to the surface parametrization within each visual area (we call this condition the *topological condition* within each visual area). Second, since the human visual cortex is organized into several visual areas, and the visual areas hold the same organization (for instance, V1 is adjacent to V2d and V2v) across subjects (to distinguish, we call this condition *topology-preserving condition* across visual areas).

Both conditions can be quantified by the Beltrami coefficient of quasiconformal maps [25]. Namely, we will use the Beltrami coefficient to enforce a topology-preserving surface segmentation and use the concept again to ensure the topological condition for the retinotopic map within each visual area. Next, we introduce the Beltrami coefficient's definition and then explain how it is adopted into these two conditions.

An orientation-preserving homeomorphism $\varphi(z): \mathbb{C} \to \mathbb{C}$ is *quasiconformal* if it satisfies the Beltrami equation $\frac{\partial \varphi(z)}{\partial \bar{z}} = \mu_\varphi \frac{\partial \varphi(z)}{\partial z}$, where $\mu_\varphi$ is the *Beltrami coefficient* satisfying $|\mu_\varphi|_\infty < 1$. In particular, if $\mu_\varphi = 0$ then $\varphi$ is conformal.

The topological condition requires the retinotopic map from each visual area to the retina to be orientation consistent. The orientation (biologically the visual field sign, *VFS*) for the visual area can be either positive or negative, depending on the parametrization and visual area. Once fixing a parametrization, the map (from parametric space to retina) orientation for a specific visual area is then fixed. Of course, one can treat the negative visual field area to be positive by flipping the parametrization vertically

($u'^{(2)} = -u^{(2)}$). Without losing generality, considering a positive visual area, we require the VFS is consistently positive. Mathematically, it requires the Beltrami coefficient $\mu_f$, associated with the retinotopic map from parametric space (within the unit disk) to visual field space $f: u_i \mapsto v_i$, satisfies $|\mu_f|_\infty < 1$.

We used a registration-akin method for topology-preserving segmentation. The method diffeomorphically morphs the topology-prior (which has the right topology) to the subject's parametrization disk and then uses prior's label information for the subject. The segmentation is topology-preserving if the morphing is diffeomorphic. Let $g: D \to D$ be the *morph function*. If the Beltrami coefficient $\mu_g$ satisfies $|\mu_g|_\infty < 1$, $g$ is diffeomorphic [18]. Compared to retinotopic registration [18], where visual coordinates are required when defining a template, we only require the prior label, which is easier to give and reduces the model visual coordinates hypothesis.

### 2.5 tRF Model

We introduce our tRF model, which includes fMRI decoding $\boldsymbol{T} = \{(\boldsymbol{v_i}, \sigma_i)|\ V_i \in V\}$ and visual areas segmentation $\boldsymbol{L} = \{L_i \in \mathbb{N}|\ V_i \in V\}$, (one value for an area) altogether.

**Topology-preserving segmentation**

We search a wrapping function $g$ that maximizes the label alignment between subject and wrapped-prior. However, there are no labels before segmentation. As an alternative, we maximize the alignment between VFS of topology prior and VFS of a subject.

Let $J = (F_J, E_J, u_J, L_J)$ be the topology-prior mesh, where $F_J$ is face set, $E_J$ is edge set, $u_J = \{u_j\}$ is vertex set and $L_J = \{L_j \in \mathbb{N}|\ u_j \in u_J\}$ is label set, $\hat{s}_g: D \to \{-1,1\}$ be the VFS for topology-prior after wrapping, and $s \coloneqq \text{sign}(1 - |\mu_f|)$ be the VFS function of the subject, where $\mu_f = \frac{\partial v}{\partial \bar{u}} / \frac{\partial v}{\partial u}$ ($v = v^{(1)} + iv^{(2)}$ and $u = u^{(1)} + iu^{(2)}$). We model the registration by $E = \int_D |\hat{s}_g - s| + \lambda_g \int_D |\nabla g|^2\, du$, such that $|\mu_g| < 1$, where $|\hat{s}_g - s|$ is the *VFS mismatch cost,* and $\lambda_g$ is a positive constant controls the smoothness of $g$. The mismatch cost is formulated to encourage the subject, and the wrapped topology-prior shares the same VFS at each position.

**Topological fMRI decoding**

Without losing generality, for a positive orientated visual area $V^+$, the topological fMRI decoding model is to solve within the $V^+$ area, by $\boldsymbol{T} = \arg\min_{\boldsymbol{T}} \sum_{V_j \in V^+} \int |\hat{y}(t; \boldsymbol{v_j}, \sigma_j) - y_j(t)|^2\, dt + \lambda_v |\nabla \boldsymbol{v}|_i^2\, du$, s.t. $|\mu_f| < 1$, where $|\nabla \boldsymbol{v}|^2 = |\nabla v^{(1)}|^2 + |\nabla v^{(2)}|^2$. Similarly, the tRF model can be applied to the negative area by flipping the parametrization vertically. To avoid over smoothing, we choose the values of $\lambda_v$ based on the Generalized Cross-Validation [26, 27].

**tRF model**

The tRF finds the visual parameters for the visual cortex via two modules,

$$\hat{\boldsymbol{L}} = L_J(g),\ \text{where}\ g = \arg\min_g \int_D |\hat{s}_g - s| + \lambda_g \int_D |\nabla g|^2\, du,\ \text{s.t.}\ |\mu_g| < 1 \quad (2)$$

$$\hat{\boldsymbol{T}} = \arg\min_{\hat{\boldsymbol{T}}} \sum_{u_i \in D} \int (\hat{y} - y_i)^2 dt + \lambda_v |\nabla \boldsymbol{v}|_i^2,\quad \text{s.t.}\quad s(u_i) = \hat{s}(u_i|\hat{\boldsymbol{L}}). \quad (3)$$

## 2.6 Numerical Method

A direct search for $(\hat{L}, \hat{T})$ is computationally infeasible since the number of parameters is typically high and $\hat{y}$ is also computationally heavy. Naturally, we divide the problem into two subproblems, segmentation and fMRI decoding, and update them iteratively. Since both subproblems constrain the norm of Beltrami, we introduce the technique in advance, which is called *topology-projection*.

**Topology-projection with smoothing**

Given a map $f$, the topology-projection computes a smooth and topological map $f'$ with small changes of $f$. According to Measurable Riemannian Mapping Theorem [25], the Beltrami coefficient uniquely encodes a map upon suitable normalization. Therefore, we can manipulate a map by its Beltrami coefficient. If a maps whose $|\mu_f| > 1$ for some points, we adjust its norm for those points by $\mu_f' = a\mu_f/|\mu_f|$ ($a = 0.95$ in this work). Then we use $\mu_f'$ to recovery the topological map $f'$. In specific, one can find $f'$ by solving $\nabla \cdot A \nabla f' = 0$, with Dirichlet boundary condition. Here A is a 2-by-2 matrix upon $\mu_f'$ [28], $\nabla$ and $\nabla \cdot$ are the gradient and divergence operators, respectively. $f'$ can be solved efficiently by writing $\nabla \cdot A \nabla$ as a matrix (see Supplementary Materials - *SM*). After the topology-projection, we then apply *Laplacian smoothing* [27] to $f'$ to make it smooth.

**Topology-preserving segmentation**

To solve **Eq.** 2, we divide it into two subproblems: naïve $g$ searching and topology-projection. Specifically, (1) for $u \in u_J$, we update the target of $g(u)$ to nearest point $g'$ such that $s(u_i) = \hat{s}(g')$, since it minimizes the VFS mismatch; (2) used the topological projection to fix the topological condition for $g'$ (see *SI* for segmentation results).

**Topological fMRI decoding**

We also split **Eq.** 3 into two subproblems: parameter searching and topology-projection. Specifically, we (1) used the topology-projection for each visual area respectively to get topological visual parameters $T'$; (2) used the gradient descent to update $T'$ parameters, $T' \leftarrow T' - \eta \left\{ \left( \frac{\partial E_p}{\partial v^{(1)}}, \frac{\partial E_p}{\partial v^{(2)}}, \frac{\partial E_p}{\partial \sigma} \right) \right\}$, where $\eta$ is a constant (*updating step*). The updated result $T'$ is further used to improve the segmentation, described in **Eq.** 2.

## 2.7 Algorithm

We now summarize the overall procedures of the tRF framework in **Alg.** 1. Note that we decay the updating step by $\eta^{(t+1)} = k_\beta \eta^{(t)}, k_\beta < 1$, to ensure tRF converges. Its convergence proof is provided in SM.

Algorithm 1: The tRF framework

---

**Input:** Discrete region of interest $S = (V, E, F)$; fMRI signal $y_i$ for each point; a tolerance $\epsilon \in \mathbb{R}^+$; and a decay factor $k_\beta < 1$.

**Results:** The topology-preserving segmentation $L$ and topological retinotopic map $T = \{(v_i, \sigma_i)\}$ that sufficiently explains the fMRI signal.

Solve the pointwise pRF parameter $(v_i, \sigma_i)$ by **Eq.** 1
Compute the conformal disk parametrization $u_i = c(V_i), V_i \in V$
Initialize $T = \{(v_i, \sigma_i)\}$, and $\delta \leftarrow \infty$
**while** $\delta > \epsilon$ **do**
    Compute the segmentation $L$, by solving **Eq.** 2
    Update $T'$ on given segmentation $L$ by solving **Eq.** 3
    $\delta = \max(T' - T)$, $T \leftarrow T'$, and $\eta^{(t+1)} \leftarrow k_\beta \eta^{(t)}$
**end**

---

## 3 Dataset

### 3.1 Synthetic data

We first evaluated our method on synthetic data. We used double-sech model [15] as ground truth and assigned a receptive field size $\sigma = k_2 v^{(1)}$ ($k_2 = 0.01, 0.02, 0.03$ for V1, V2, and V3, respectively). Then we generated the normalized fMRI signal $y_0(t)$ with the specific stimulation pattern in [29] and added two levels of noise to the fMRI signal $y(t) = y_0(t) + r(t)$, where $r \sim N(0, \gamma)$ ($\gamma = 0.1$ and $\gamma = 0.5$ respectively).

### 3.2 Real data with 7T MRI system

We also tested the framework on the Human Connectome Project (HCP) dataset [30]. The HCP dataset is a publicly available retinotopy dataset on 7T fMRI scanners with high spatial and temporal resolutions.

## 4 Results

### 4.1 Results on synthetic data

We compared the performance of the proposed method with several methods, including the pRF model, the model fitting result [18], the Laplacian smoothing, and the registration method (which registers pRF to another parameterized Double-Sech template and uses the template's value) [31]. More specifically, given the noisy functional signal $y(t)$ for each vertex, we compared the errors between methods' output and ground truth in **Tab.** 1, including the perception center error $|\Delta v|$, receptive field size error $|\Delta \sigma|$, and the numbers of triangles that violates the required VFS within its segmentation.

    One can see, (1) only the proposed method can make topological results ($T_n = \mathbf{0}$), and the smoothing achieved the worst results, which means the smoothing does not contribute to our topological results in topology-projection; (2) On the other hand, the smoothing method also achieved decent precision ($|\Delta v| \sim 2.8$); (3) The model-fitting method is promising in topological condition ($T_n < 5$). However, it is only applicable to the V1-V3 complex but not to other visual areas (see **Tab.** 2 the flipping number is significant due to regions beyond V3); (4) The TPS registration method can ensure the topology in theory. However, since the noisy measure of retinotopic coordinates, the

anchors/landmarks are noisy and destroyed the topological condition. Those results suggest the proposed method achieve the best accuracy under topological condition. We now provide an illustrative comparison in **Fig.** 3 for the mentioned methods.

| Method | $|\Delta v|$ | $|\Delta \sigma|$ | $T_n$ |
|---|---|---|---|
| pRF Method | 2.924/2.313 | 0.902/1.100 | 393/483 |
| Smoothing | 2.827/2.288 | 0.863/1.021 | 469/528 |
| Registration | 3.549/3.518 | 0.974/0.971 | 440/444 |
| Model-fitting | 3.559/3.590 | 0.902/1.100 | 3/5 |
| tRF (proposed) | **2.485/1.801** | **0.768/0.857** | **0/0** |

**Table 1.** The performance compare (the smaller, the better) on synthetic data. Results for $\gamma = 0.1$ (small noise) and $\gamma = 0.5$ (big noise) are separated by the "/" symbol.

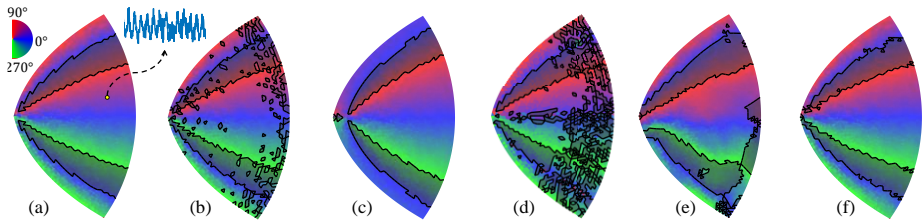

**Fig. 3.** Comparison of various methods for the V1-V3 model: (a) Ground-truth, (b) pRF, (c) model-fitting, (d) smoothing, (e) registration, and (f) the proposed tRF.

### 4.2 Results on the real data

We applied the framework to subjects in the HCP dataset. Since no ground-truth is available for the real dataset, we compared the methods based on the fMRI fitting goodness (the RMSE of fitting error $\bar{e}$) and the number of non-topology triangles ($T_n$). We list the fitting error $\bar{e}$ and topology violations $T_n$ for the first three observers in **Tab.** 2. The proposed method achieved the best fMRI fitting and zero topology violations. We also showed the intuitive comparison between the methods for the first observer in **Figs.** 4a-e. The proposed tRF (**Fig.** 4e) fixed the topology violations ($T_n = 0$) compared to the pRF model (**Fig.** 4a, $T_n = 870$), model-fitting (**Fig.** 4b, $T_n= 489$), smoothing (**Fig.** 4c, $T_n = 842$), or registration (**Fig.** 4d, $T_n = 953$). Besides, our method achieved the smallest fMRI fitting error (RMSE is **0.296**), compared to pRF (0.300), model-fitting (0.335), smoothing (0.301), or registration method (0.305). We further report the average results for the first twenty observers in the dataset. Our method reduced the mean fitting error from the second-best method's 0.376 (pRF method) to 0.372, and the topology violations were reduced from 1234 (median value) to **0**. Those results showed that the proposed tRF not only reduced the topology violation (i.e., compatible with neurophysiology conclusions) but also improved/kept the fitting power.

| Observers | pRF | Model-Fitting | Smoothing | Registration | tRF(proposed) |
|---|---|---|---|---|---|
| S1 | 0.300/870 | 0.335/489 | 0.301/842 | 0.305/953 | **0.296/0** |
| S2 | 0.252/1119 | 0.288/934 | 0.253/1126 | 0.287/1141 | **0.250/0** |
| S3 | 0.276/1194 | 0.296/808 | 0.276/1197 | 0.298/1082 | **0.273/0** |

**Table 2.** The $\bar{e}/T_n$ (fitness/topology) of different methods for the first three HCP observers.

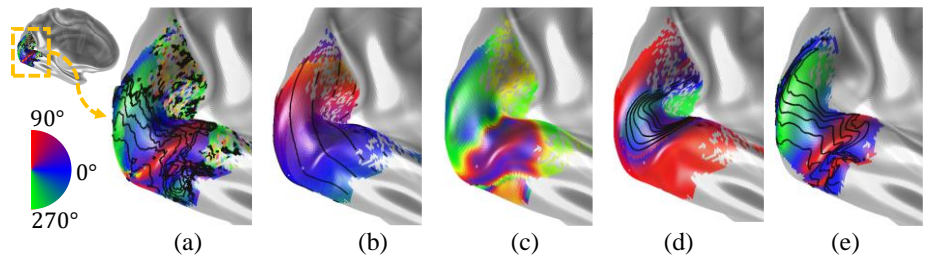

**Fig. 4.** Results of (a) pRF, (b) model-fitting, (c) smoothing, (d) registration, and (e) tRF.

# Supplementary Information for "Topological Receptive Field Model for Human Retinotopic Mapping"

## 1     Definition of $\nabla \cdot A\nabla$

$$(\nabla \cdot A\nabla)_{i,j} = \begin{cases} \sum_{[u_i,u_j,u_k]\in N(i)} \dfrac{(As_j)\cdot s_i}{|[u_i, u_j, u_k]|}, & \text{if } i \neq j \\ -\sum_{k\neq i} L_{i,k}, & \text{if } i = j \\ 0, & \text{otherwise,} \end{cases}$$

where $N(i)$ is the set of triangles attached to vertex $i$, $A = \begin{pmatrix} \alpha_1 & \alpha_2 \\ \alpha_2 & \alpha_3 \end{pmatrix}$ is a matrix with values $\alpha_1 = \dfrac{(\rho-1)^2+\tau^2}{\rho^2+\tau^2-1}$, $\alpha_2 = \dfrac{-2\tau}{\rho^2+\tau^2-1}$, and $\alpha_3 = \dfrac{(\rho+1)^2+\tau^2}{\rho^2+\tau^2-1}$, $s_i = n \times (u_j - u_k)$ denotes a vector that is perpendicular to edge $u_j - u_k$ and face normal $n$ with its norm equals $|u_j - u_k|$. Similarly, $s_j = n \times (u_k - u_i)$ and $s_k = n \times (u_i - u_j)$.

## 2     Results of topology-preserving segmentation

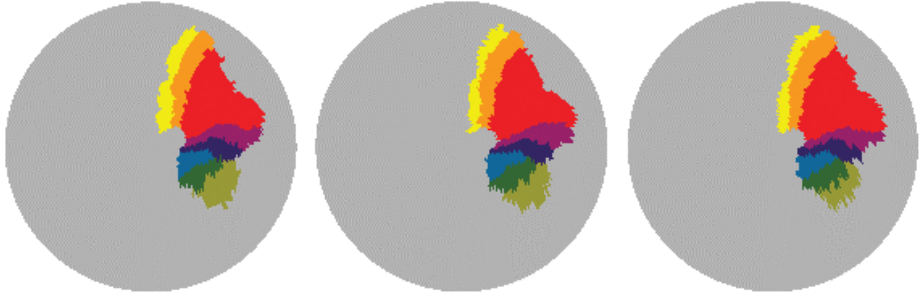

**Fig. 5S.** Segmentation results for the first three subjects

## 3     Proof for the convergence of tRF

*Proof:* Recall the fMRI decoding includes two steps: topology-projection and gradient descent. It is not hard to see, $\lim_{t\to\infty} \eta^{(t)}\nabla E_p = 0$, since $\lim_{t\to\infty} \eta^{(t)} = 0$, when $k_\beta < 1$. Namely, after sufficient iterations, the fMRI decoding will not update the visual coordinates but only fixing the topological condition. Therefore, the fMRI decoding will have the orientation (for each visual field) consistent (the effect of topology-projection). On the other hand, our topology-segmentation cannot detect any point whose orientation is different from the topology-prior, therefore, the segmentation will not change $g$ anymore. In summary, after sufficient iterations, the algorithm does not change the segmentation or visual parameters. Therefore, it converges.